\documentclass[11pt]{article}
\pdfoutput=1

\usepackage{jheppub}
\usepackage{url,comment}
\usepackage{times}
\usepackage{latexsym}
\usepackage{graphicx, graphics, hyperref, amsmath, amssymb, slashed, xcolor, bbm,bm,amsthm, array}
 \usepackage{subfigure}
 \usepackage{listings} 
 \usepackage{trimclip}

\usepackage{rotating}
\usepackage{afterpage}
 
\def\beq{\begin{equation}}
\def\eeq{\end{equation}}
\def\bea{\begin{eqnarray}}
\def\eea{\end{eqnarray}}

\def\lsim{\mathrel{\rlap{\lower4pt\hbox{\hskip1pt$\sim$}}
     \raise1pt\hbox{$<$}}}        
\def\gsim{\mathrel{\rlap{\lower4pt\hbox{\hskip1pt$\sim$}}
     \raise1pt\hbox{$>$}}}

\graphicspath{{plots/}}

\title{Displaced Lepton Jet Signatures from Self-Interacting Dark Matter Bound States}

\author[a]{Yuhsin Tsai,}
\author[b,c]{Tao Xu,}
\author[b]{Hai-Bo Yu}

\affiliation[a]{Maryland Center for Fundamental Physics, Department of Physics,
University of Maryland, Stadium Dr., College Park, MD 20742-4111, USA}
\affiliation[b]{Department of Physics and Astronomy, University of California, 900 University Avenue, Riverside, CA 92521, USA}
\affiliation[c]{Zhejiang Institute of Modern Physics, Department of Physics, Zhejiang University, 38 Zheda Road, Hangzhou, Zhejiang 310027, China}

\emailAdd{yhtsai@umd.edu}
\emailAdd{taoxu@zju.edu.cn}
\emailAdd{haiboyu@ucr.edu}

\abstract{

We study self-interacting dark matter signatures at the Large Hadron Collider. A light dark photon, mediating dark matter self-interactions, can bind dark matter particles to form a bound state when they are produced via a heavy pseduoscalar in $pp$ collisions. The bound state can further annihilate into a pair of boosted dark photons, which subsequently decay into charged leptons through a kinetic mixing portal, resulting in striking displaced lepton jet signals. After adapting the analysis used in the ATLAS experiment, we explore the reach of the model parameters at the $13~{\rm TeV}$ run with an integrated luminosity of $300~{\rm fb^{-1}}$. For heavy dark matter, the displaced lepton jet searches can surpass traditional monojet signals in setting the lower bound on the pseduoscalar mass. If a positive signal is detected, we can probe the dark matter mass and the dark coupling constant after combining both the displaced lepton jet and monojet searches. We further show the CMS dimuon search can be sensitive to the final state radiation of the dark photon. Our results demonstrate terrestrial collider experiments complement astronomical observations of galaxies in the search of the self-interacting nature of dark matter. }

\begin{document}

\begin{flushright}
\small{.}
\end{flushright}

\maketitle

\section{Introduction}

Astrophysical observations indicate that the dark matter sector could be more complex and vibrant than we thought. For example, the rotation curves of spiral galaxies exhibit a great diversity~\cite{Moore:1994yx,Flores:1994gz,Persic:1995ru,deBlok:2001hbg,deNaray:2009xj,Oman:2015xda}, which is hard to understand in the vanilla cold dark matter theory. A detailed study of mass distributions in galaxy clusters reveals that there are dark matter density cores in their inner regions~\cite{Newman:2012nw}, in contrast to cusps predicted in cold dark matter~\cite{Dubinski:1991bm,Navarro:1995iw}. It is not clear whether the cold dark matter model with baryonic feedback can reconcile these small-scale discrepancies~\cite{Oman:2015xda,Santos-Santos:2017,Schaller:2014gwa}. On the other hand, it has been shown that they can be resolved if dark matter has strong self-interactions, analogous to the nuclear interactions, see~\cite{Tulin:2017ara} for a review and references therein and~\cite{Kaplinghat:2015aga,Kamada:2016euw,Creasey:2016jaq,Valli:2017ktb,Ren:2018jpt} for detailed fits to observational data. Moreover, taking astrophysical observations over different scales from dwarf galaxies to galaxy clusters, we may probe dark matter particle physics parameters~\cite{Kaplinghat:2015aga} and even production mechanisms~\cite{Huo:2017vef}.

Aside from these astro colliders probing dark matter self-interactions, there are terrestrial particle colliders such as the Large Hadron Collider (LHC) that can test dark matter interactions with the standard model (SM) particles. For example, the missing energy signal, which comes from seeing an imbalance of visible particle momentum due to the existence of invisible dark matter particles, provides an important tool for dark matter hunting at the LHC~\cite{Sirunyan:2018dsf,ATLAS:2018vvx}. However, the missing energy search does not directly probe the interactions in the dark sector. Given the compelling astrophysical hints for strong dark matter-dark matter interactions, we explore complementary signatures to test the self-interacting nature of dark matter at the LHC.

\begin{figure}
\begin{center}
\includegraphics[width=8cm,height=3.5cm]{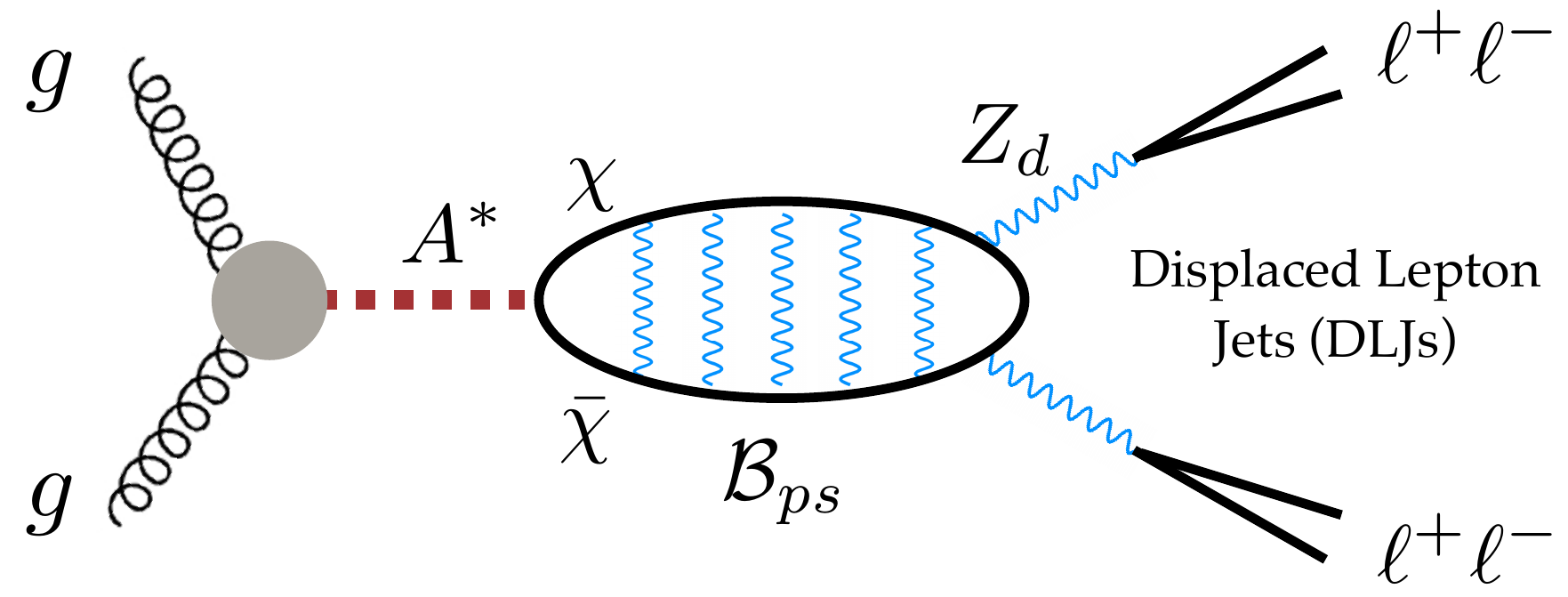}
\caption{Displaced lepton jet signatures from the SIDM bound state at the LHC. In the model we consider, a heavy pseduoscalar ($A$) couples the SIDM particle ($\chi$) to gluons ($g$), a dark photon ($Z_d$) mediates dark matter self-interactions and leads to formation of the bound state ($\mathcal{B}_{ps}$). The boosted $Z_d$ decays to SM charged leptons via a kinetic mixing portal.}
\end{center}
\label{fig:feydigm}
\end{figure}

In many particle physics models of self-interacting dark matter (SIDM)~\cite{Feng:2009hw,Buckley:2009in,Loeb:2010gj,Frandsen:2011kt,Aarssen:2012fx,Tulin:2012wi,Tulin:2013teo,Cline:2013pca,Kahlhoefer:2013dca,Laha:2013gva,Foot:2014uba,Boddy:2014yra,Bernal:2015ova,Laha:2015yoa,Cyr-Racine:2015ihg,Boddy:2016bbu,Ma:2016tpf,Baldes:2017gzw,Kahlhoefer:2017umn,Baldes:2017gzu,Kamada:2018zxi,Braaten:2018xuw,Duerr:2018mbd,Essig:2018pzq,Kamada:2018kmi}, there exists a dark force mediator that is much lighter than the dark matter particle (but see~\cite{Chu:2018fzy}). In this case, when a pair of SIDM particles is produced at the LHC, they may form a bound state due to the same mediator that leads to dark matter self-interactions in the halos. The resulting bound state can annihilate into two boosted mediators, which subsequently decay back to the SM particles, as illustrated in Fig.~\ref{fig:feydigm}. If the mediator's coupling to the SM is small enough to satisfy other existing constraints, it can be long-lived and have a macroscopic decay length comparable to the size of the LHC detectors. Compared to prompt signals, the long-lived particle search has much less SM backgrounds. And the resonance feature can further help us distinguish the signal from the backgrounds that are combinatorial and typically monotonic in energy. Thus, searching for the mediator from the bound state decay provides a powerful way of testing SIDM models.

In this paper, we propose an LHC study of SIDM using the production of the dark matter bound states and their decay into displaced lepton jets (DLJs). As we will show, the DLJ search at the Run 3 LHC can be sensitive to the SIDM parameters that resolve the discrepancies on galactic scales. The reconstruction of the bound state mass through the DLJ energy gives a measurement of the dark matter mass, and a comparison between the bound state and missing energy signatures provides information about dark matter self-interactions, complementary to astrophysical observations. Note the idea of looking for dark matter bound states at particle colliders has been discussed before~\cite{Shepherd:2009sa,An:2015pva,Tsai:2015ugz,Bi:2016gca,Li:2017xyf,Elor:2018xku,Krovi:2018fdr}. Here we focus on the DLJ signatures from the long-lived dark force mediator in the SIDM context. A similar analysis was carried out using B-factory~\cite{An:2015pva} and LHC~\cite{Tsai:2015ugz} results. In this work, we study a simplified SIDM model with a heavy pseudoscalar that couples SM quarks to dark matter particles. We further take the advantage of non-conventional ATLAS triggers searching for long-lived particles~\cite{ATLAS:2016jza} and conduct a detailed study of the DLJ signals from the SIDM bound states at the LHC Run 3. We also explore the possibility of narrowing down the dark matter model parameters after combing monojet, multi-muon and DLJ searches, as well as astrophysical observations. 

The outline of this paper is as follows. In Sec.~\ref{sec:darkphoton}, we present the particle physics model and discuss various constraints on model parameters. In Sec.~\ref{sec:DLJs}, we discuss DLJ searches of the SIDM bound state. In Sec.~\ref{sec:dimuon}, we discuss prompt multi-muon searches, especially for the final state radiation of the dark photon. In Sec.~\ref{sec:results}, we show current LHC bounds, projected future reaches and their implications for astrophysical observations. We conclude in Sec.~\ref{sec:con}.

\section{The SIDM model}
\label{sec:darkphoton}

We consider an SIDM scenario, where a fermionic dark matter particle ($\chi$) couples to a dark photon ($Z_d$) with mass $m_{Z_d}$ and kinetic mixing $\epsilon_{Z_d}$ to the SM photon~\cite{Holdom:1985ag,Foot:2004pq,Pospelov:2007mp,ArkaniHamed:2008qn}, 
\begin{equation}
\label{eq:SIDM}
\mathcal{L}_{\rm SIDM}= \bar{\chi}\left(i\slashed\partial+g_{\chi}\slashed Z_d-m_{\chi}\right)\chi+\frac{m_{Z_d}^2}{2}Z_{d,\mu}Z_d^{\mu}+\frac{\epsilon_{Z_d}}{2} F_{\mu\nu}F^{\mu\nu}_d.
\end{equation}
The dark photon $Z_d$ mediates dark matter self-interactions and bind dark matter particles into a bound state. The SIDM bound state decays dominantly into two dark photons, which can further decay into SM leptons via the kinetic mixing term. The decay length of $Z_d$ in the lab frame is
\begin{eqnarray}\label{eq:Zddecay}
L_{Z_d\to \ell^+\ell^-}&=& \gamma_{Z_d}\left[\sum_\ell \frac{\alpha\,\epsilon_{Z_d}^2m_{Z_d}}{3}\sqrt{1-\frac{4m_{\ell}^2}{m_{Z_d}^2}}\left(1+\frac{2m_{\ell}^2}{m_{Z_d}^2}\right)\right]^{-1}
\\
&\sim&1\,{\rm m}\,\left(\frac{m_{\chi}}{10\,{\rm GeV}}\right)\left(\frac{10\,{\rm MeV}}{m_{Z_d}}\right)^2\left(\frac{10^{-4}}{\epsilon_{Z_d}}\right)^2, 
\end{eqnarray}
where $\gamma_{Z_d}\approx m_{\chi}/m_{Z_d}$ is the boost factor.
 
\begin{figure*}
\begin{center}
\includegraphics[width=7.45cm]{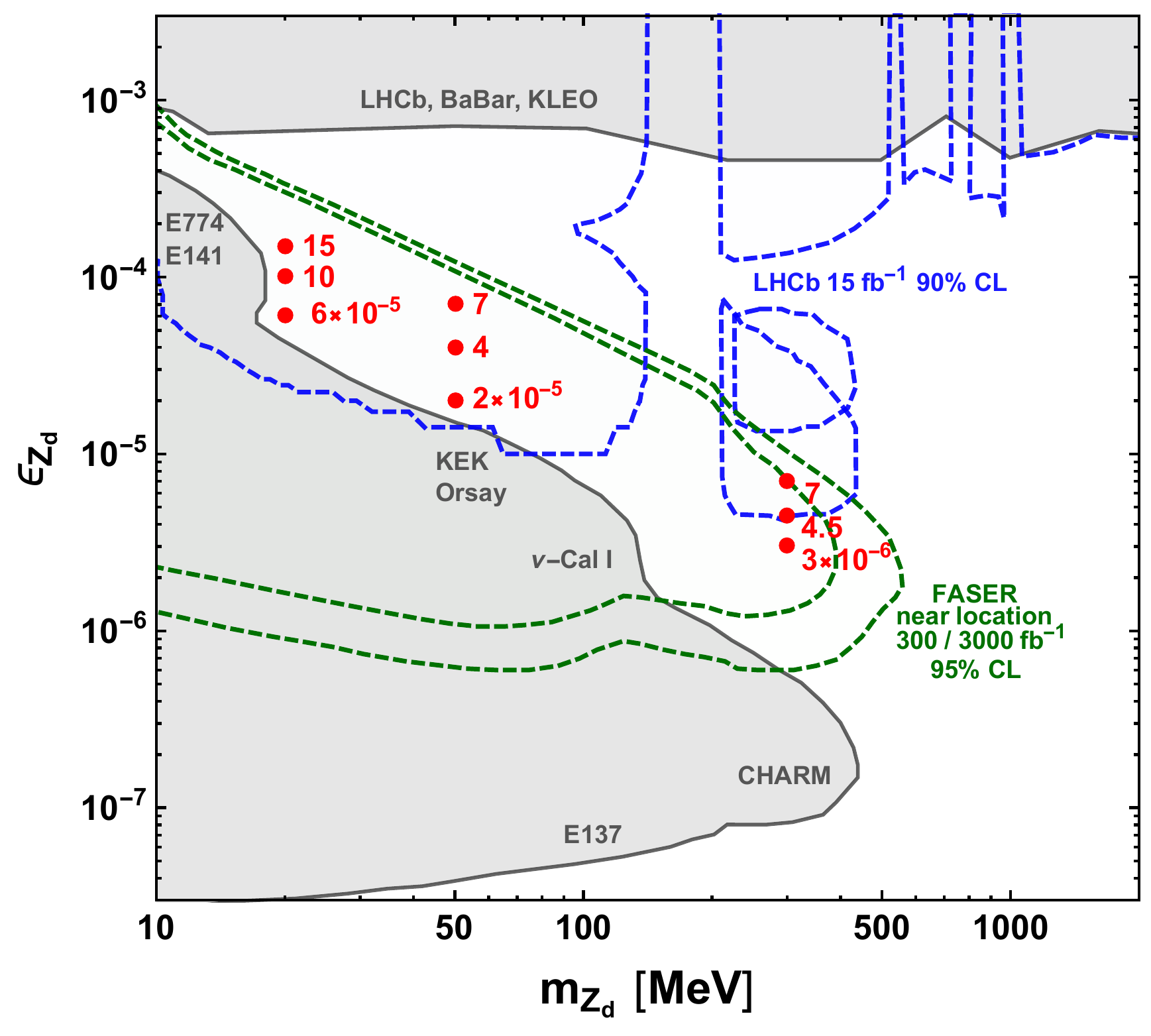}\quad\includegraphics[width=7.5cm]{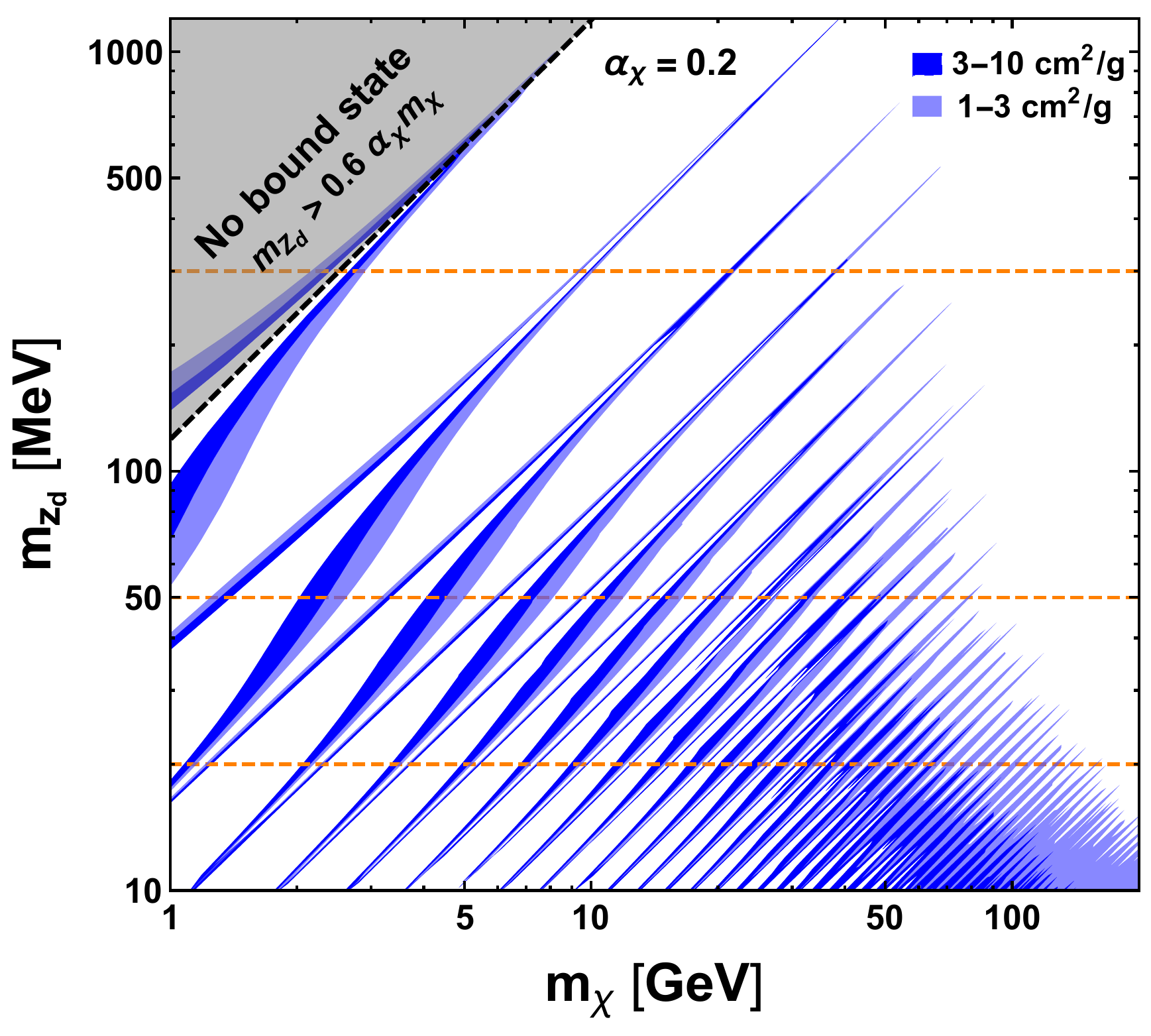}
\end{center}
\caption{\emph{Left:} Dark photon masses and kinetic mixing parameters that we use for the collider study (red dots). We also show the existing bounds from the BaBar~\cite{Lees:2014xha}, LHCb~\cite{Aaij:2017rft} and beam dump experiments (gray shaded), see, e.g.,~\cite{Bjorken:2009mm,Andreas:2012mt,Battaglieri:2017aum}. Future measurements from the LHCb~\cite{Ilten:2016tkc} (blue dashed) and the proposed FASER experiment~\cite{Feng:2017uoz} (green dashed) can further test the model parameters. \emph{Right:} $m_{\chi}\textup{--}m_{Z_d}$ parameter regions (blue shaded), where the self-scattering cross section per mass is $\sigma/m_\chi=1\textup{--}10~{\rm cm^2/g}$ favored by astrophysical observations in dwarf galaxies. The region below the dashed black curve satisfies the bound state formation condition given in Eq.~(\ref{eq:condition}). The three orange horizontal lines denote the mediator masses we take.}
\label{fig:SIDMPara}
\end{figure*}

We take $m_{Z_d}=20~{\rm MeV}$, $50~{\rm MeV}$ and $300~{\rm MeV}$ in our collider study and for each $m_{Z_d}$ we choose three benchmark values of $\epsilon_{Z_d}$, as shown in Fig.~\ref{fig:SIDMPara} (red points). The choice of $\epsilon_{Z_d}$ values is motivated by the displaced signal search in the LHC hadronic calorimeter (HCAL) and the early part of the muon spectrometer, corresponding to $L_{Z_d\to \ell\bar{\ell}}\approx 2\textup{--}6~{\rm m}$. Since $Z_d$ could still decay in the inner part of the detector, we also include the prompt muon search in our study for the case of $m_{Z_d}=300~{\rm MeV}$.

Since we will be considering a large dark coupling constant $\alpha_{\chi}\equiv g_{\chi}^2/4\pi > 0.1$, the dark matter relic density needs to come from asymmetric dark matter scenarios where only $\chi$ particles are left in the present universe, see, e.g.,~\cite{Davoudiasl:2012uw,Petraki:2013wwa,Zurek:2013wia}. In collider experiments, however, dark matter particles are produced in $\chi\bar{\chi}$ pairs. For $\chi$ and $\bar{\chi}$ to form a bound state, the corresponding Compton wavelength of $Z_d$ should be much larger than the size of the bound state~\cite{PhysRevA.1.1577}, which requires
\begin{equation}
\label{eq:condition}
m_{Z_d}\lsim 0.6\,m_{\chi}\alpha_{\chi}. 
\end{equation}
Since for SIDM there is a rough scaling relation between the required mediator versus dark matter masses
\begin{equation}
m_{Z_d}\sim10^{-2}\textup{--}10^{-4}m_\chi,
\end{equation}
Eq.~(\ref{eq:condition}) is satisfied if $\alpha_\chi\gtrsim10^{-2}\textup{--}10^{-3}$. This means the SIDM particles can easily form a bound state when being produced near the mass threshold in colliders. 

There are a number of other constraints on the model we need to consider before discussing the collider search. Dark matter direct detection experiments can put strong upper limits on the kinetic mixing parameter~\cite{Kaplinghat:2013yxa,DelNobile:2015uua,Kahlhoefer:2017umn,Kahlhoefer:2017ddj}. A recent PandaX-II analysis shows $\epsilon_{Z_d}\lesssim10^{-10}$ if dark matter-nucleus scattering can occur via $Z_d$ exchange~\cite{Ren:2018gyx}. Within this limit, the decay length of $Z_d$ will be too long to produce DLJ signatures in the LHC detectors. To avoid the direct detection constraints, we assume that the SIDM fermions carry a small Majorana mass that breaks the dark symmetry U$(1)_d$~\cite{Schutz:2014nka,Blennow:2016gde,Zhang:2016dck}. In the mass basis, $Z_d$ couples simultaneously to the heavy and light dark matter particles, where the mass splitting comes from the Majorana mass. The resulting inelastic scattering $\chi_{\rm light}N\to\chi_{\rm heavy}N$ at direct detection experiments is kinetically forbidden by the mass difference as long as the Majorana mass is larger than the kinetic energy $\sim m_{\chi}v_{max}^2/2$ in the Milky Way halo, where $v_{max}\approx750~{\rm km/s}$ is the maximal dark matter velocity with respect to the target nucleus. For $m_{\chi}\sim100~{\rm GeV}$, the required mass splitting is only $\sim 100~{\rm keV}$. Since it is much smaller than the binding energy of the bound state ($\gsim~{\rm GeV}$), the presence of the Majorana mass has negligible effects for the collider study. 

In the presence of the mass splitting, dark matter self-scattering between the two light states is attractive~\cite{Zhang:2016dck,Blennow:2016gde}. In this case, a detailed calculation of the self-scattering cross section per mass, $\sigma/m_\chi$, is complicated and numerically expensive, which is beyond the scope of this paper. In Fig.~\ref{fig:SIDMPara} (right), we plot the parameter space (blue shaded) that gives $\sigma/m_\chi=1\textup{--}3~{\rm cm^2/g}$ (light) and $3\textup{--}10~{\rm cm^2/g}$ (dark), assuming an attractive Yukawa potential without the mass splitting. The numerical methods developed in~\cite{Tulin:2012wi,Tulin:2013teo} have been used to calculate the transfer cross section, and we take a characteristic dark matter relative velocity $v_{\rm rel}=50~{\rm km/s}$ for dwarf galaxies. For simplicity, we do not perform the thermal average of the cross section over the velocity distribution as in~\cite{Tulin:2013teo}. Including effects from the velocity averaging would slightly broaden the allowed SIDM parameter space shown in Fig.~\ref{fig:SIDMPara} (right) and smooth the resonance peaks for fixed $\sigma/m_\chi$. In addition, we also expect the favored SIDM region would shift towards lower $m_{Z_d}$ values if we include the mass splitting~\cite{Zhang:2016dck,Blennow:2016gde}, but the overall resonant features remain. This is because for given $m_\chi$ a lighter $m_{Z_d}$ is required to compensate the suppression effect caused by the mass splitting. Refs.~\cite{Kamada:2016euw,Ren:2018jpt} take $\sigma/m_\chi=3~{\rm cm^2/g}$ and show that SIDM can explain diverse galaxy rotation curves. And they also argue a wide range of $\sigma/m_\chi$ may work as well as long as it is larger than $\sim1~{\rm cm^2/g}$, due to the degeneracy effect between the cross section and halo parameters in the fits. Given the degeneracy and the expected parameter shifts in the presence of the mass splitting and velocity averaging, we consider a range of $\sigma/m_\chi$ values in Fig.~\ref{fig:SIDMPara} (right) instead of fixing it to a specific number.\footnote{Another concern is that the Majorana mass term may allow dark matter particles to annihilate through $\chi_{\rm light}\,\chi_{\rm light}\to 2Z_d$, since there can be no conserved charges that forbid the annihilation as in asymmetric dark matter scenarios. If this happens, SIDM cannot keep the observed abundance for the size of $\alpha_{\chi}$ we consider. Moreover, annihilation in halos can be boosted due to the presence of the light mediator, and there are strong constrains from indirect detection experiments for this type of models~\cite{Kaplinghat:2015gha,An:2016gad,An:2016kie,Bringmann:2016din,Cirelli:2016rnw}. However, we can easily avoid these problems by considering a more general setup, where dark fermions carry more than one flavor and charge. The Majorana mass can break only U$(1)_d$ but not all the other symmetries. In this case, $Z_d$ still couple to the heavy and light fermions simultaneously, but the fermions carry conserved charges that preserve the dark matter asymmetry. In this paper, we will focus on the collider signature of SIDM and leave detailed model building for future work.}

\begin{figure}
\begin{center}
\includegraphics[width=7.45cm]{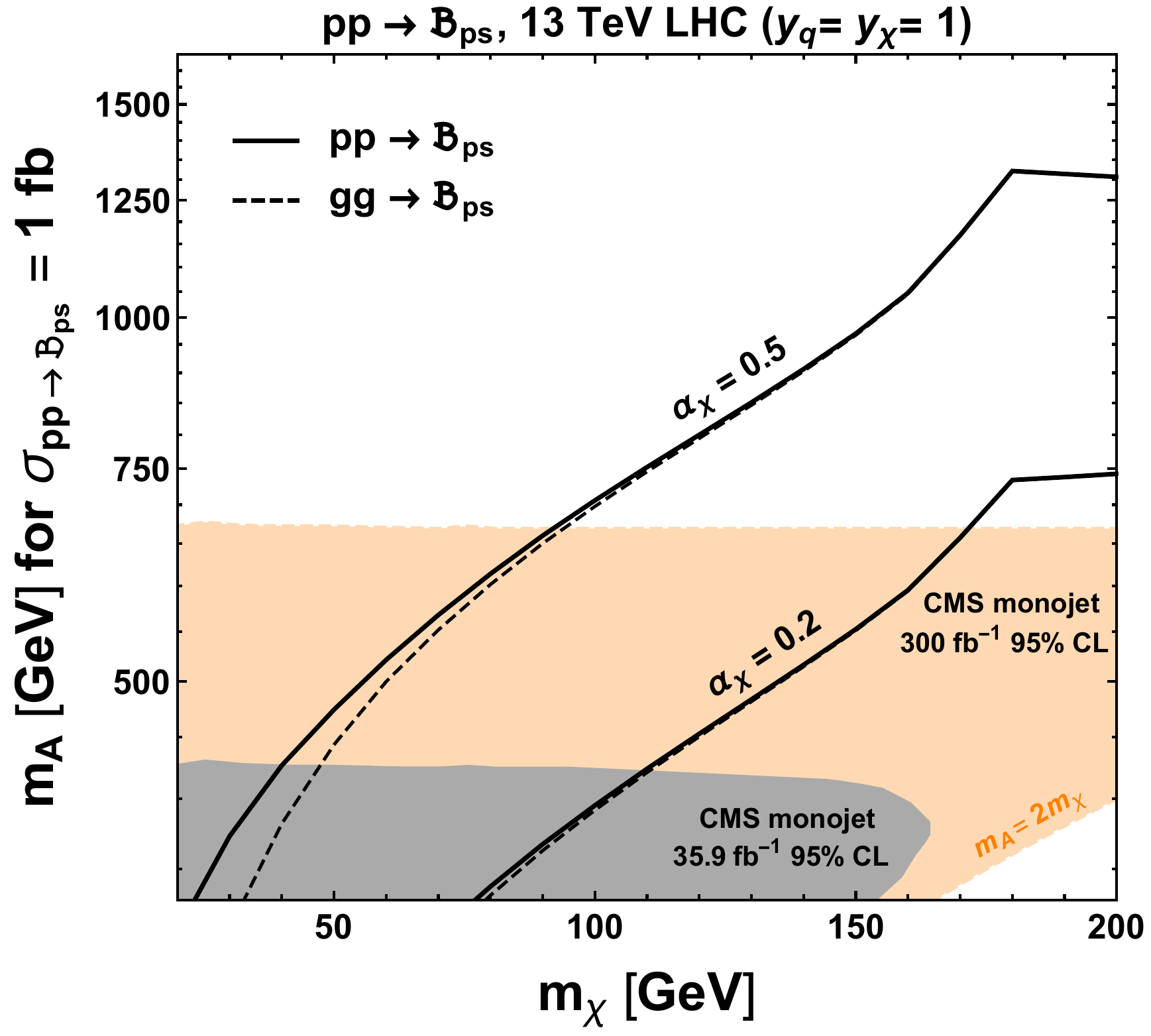}
\caption{The pseudoscalar mass $m_A$ vs. dark matter mass $m_\chi$ for having $1~{\rm fb}$ production cross section of the bound state at 13 TeV LHC, assuming $\alpha_{\chi}=0.2$ and $0.5$. The dashed (solid) curves denote the contribution from the gluon fusion process (plus the quark-initiated process). We assume the pseudoscalar couplings $y_q=y_{\chi}=1$ and directly take the $2\sigma$ exclusion limit on $m_A$ from the CMS monojet search~\cite{Sirunyan:2017jix} (gray shaded). A projected monojet bound with an integrated luminosity of $300~{\rm fb^{-1}}$ is also shown (orange shaded).}
\label{fig:productionrate}
\end{center}
\end{figure}

For the collider study, we focus on the $s$-wave production of a pseudo-scalar bound state $\mathcal{B}_{ps}$ shown in Fig.~\ref{fig:feydigm}. In particular, we consider a heavy pseudoscalar $A$ that couples the dark matter particle to the SM~\cite{Berlin:2015wwa,Banerjee:2017wxi},
\begin{equation}
\mathcal{L}_{A}\supset\frac{1}{2}(\partial A)^2-\frac{m_A^2}{2}A^2-iy_{\chi}A\bar{\chi}\gamma_5\chi-i\,\frac{y_q}{\sqrt{2}}\,\displaystyle{\sum_f}\frac{m_{f}}{v}A\bar{f}\gamma_5f,
\end{equation}
where $v=174~{\rm GeV}$ and $f$ represents the SM fermions. Although $A$ can lead to dark matter-nucleus scattering, direct detection constraints on $m_A$ are very weak because the cross section is either highly momentum or loop suppressed~\cite{Haisch:2013uaa,Li:2018qip}. The pseudoscalar can be produced at the LHC through heavy-quark and gluon-fusion processes. For the mass and coupling we consider, the best existing constraint on $m_A$ comes from the CMS monojet search~\cite{Sirunyan:2017jix}. It applies for $m_A>2m_{\chi}$ when the missing energy decay $A\to\chi\bar{\chi}$ dominates the branching ratio. Using the Collider Reach tool~\cite{collreach} to rescale the bound based on the parton distribution function (PDF) and the luminosity increase, we estimate the $m_A$ reach with $300$ fb$^{-1}$ of data (orange) and show it together with the current constraint (gray) in Fig.~\ref{fig:productionrate}.

Following the calculation of the Yukawa bound state~\cite{An:2015pva,Tsai:2015ugz}, the production cross section of the SIDM bound state from the quarks is given by 
\begin{eqnarray}
\sigma_{q\bar{q}\to\mathcal{B}_{ps}}=\frac{\zeta(3)\,\alpha_{\chi}^3\,y_{\chi}^2\,y_q^2\,m_{\mathcal{B}}^4}{96\,s\,\left[(m_{\mathcal{B}}^2-m_{A}^2)^2+(\Gamma_A\,m_A)^2\right]}
\times\left[\displaystyle{\sum_q}\,\frac{m_q^2}{v^2}\displaystyle{\int^1_{m_{\mathcal{B}}^2/s}}\frac{dx}{x}f_q(x)f_{\bar{q}}(\frac{m_{\mathcal{B}}^2}{x\,s})+(q\leftrightarrow\bar{q})\right],\nonumber
\end{eqnarray}
where $f_{q,\bar{q}}(x)$ is the PDF and $\sqrt{s}=13$ TeV for the LHC search. The factor $\alpha_{\chi}^3$ comes from the wavefunction square of dark matter particles at zero separation $|\psi(0)|^2=\alpha_{\chi}^3m_{\chi}^3/8\pi$. The bound state and mediator masses come from the eigenvalues of the bare mass matrix~\cite{Elor:2018xku}
\begin{equation}
\begin{pmatrix}
m_{A,0}^2-i\,m_{A,0}\,\Gamma_{A,0} & -m_{A\mathcal{B}_{ps}}^2 \\
-m_{A\mathcal{B}_{ps}}^2 & m_{\mathcal{B},0}^2 -i\,m_{\mathcal{B},0}\,\Gamma_{\mathcal{B}_{ps},0}
\end{pmatrix}.
\end{equation}
The bound state mass before the mixing $m_{\mathcal{B},0}=2m_{\chi}-E_b$, where $E_b=m_{\chi}\alpha_{\chi}^2/4$ is the ground state binding energy. The $A$-$\mathcal{B}_{ps}$ mixing $m_{A\mathcal{B}_{ps}}^2\approx y_{\chi}^2\alpha_{\chi}^3m_{\chi}^2/8\pi$. Since $m_{A\mathcal{B}_{ps}}^2\ll m_{\mathcal{B},0}^2$ for the $\alpha_{\chi}$ we consider, the physical bound state mass $m_{\mathcal{B}}\approx m_{\mathcal{B},0}$. The width of the pseudoscalar is dominated by the decay into dark matter particles $\Gamma_{A,0}\sim y_{\chi}^2m_{A,0}/8\pi$. The bound state width is estimated to be $\Gamma_{\mathcal{B},0}\sim \alpha_{\chi}^5m_{\chi}$ from the bound state decay into two $Z_d$'s with transverse polarization. For the mass and coupling we consider, the decay rate into dark photons is much larger than the decay into SM fermions $\Gamma_{\mathcal{B}\to f\bar{f}}\sim y_{\chi}^2y_q^2\alpha_{\chi}^3m_f^2m_{\chi}^5/(v^2m_A^4)$.

The production rate from the gluon fusion process is given by 
\begin{eqnarray}\label{eq:Bprod}
\sigma_{gg\to\mathcal{B}_{ps}}=\frac{\pi\zeta(3)\,\alpha_{\chi}^3\,y_{\chi}^2\,m_{\mathcal{B}}^6}{256\,s\,\Lambda_1(m^2_{\mathcal{B}})^2\left[(m_{\mathcal{B}}^2-m_{A}^2)^2+(\Gamma_A\,m_A)^2\right]}
\times\left[\displaystyle{\int^1_{m_{\mathcal{B}}^2/s}}\frac{dx}{x}f_g(x)f_{g}(\frac{m_{\mathcal{B}}^2}{x\,s})\right].
\end{eqnarray}
The one loop gluon-fusion coupling is given by~\cite{Gunion:1989we,Gu:2015lxj}
\begin{eqnarray}
\Lambda_1(q^2)^{-1}&=&\frac{\alpha_s\,y_q}{16\sqrt{2}\pi\,v}\displaystyle{\sum_{q=t,b,...}}
A_{1/2}^{A}(\tau_q),
\quad \\A_{1/2}^{A}(\tau)
&=&2\,\tau^{-1}\times\begin{cases}
[\sin^{-1}(\sqrt{\tau})]^2,\quad\tau\leq 1\\
-\frac{1}{4}\left[\log\left(\frac{1+\sqrt{1-\tau^{-1}}}{1-\sqrt{1-\tau^{-1}}}-i\,\pi\right)\right]^2,\,\,\tau>
1,
\end{cases}
\end{eqnarray}
where $\tau_q \equiv{q^2}/{4m_q^2}$ and $\alpha_s\approx 0.12$ is the SM QCD coupling. 

We also include the $K$-factor for both the gluon fusion and quark-initiated processes when calculating the cross section for collider production. The values for the gluon fusion process are $3.2$ ($2.3$) for $m_{\mathcal{B}}=100$ ($400$) GeV~\cite{Ahmed:2016otz}. For the quark process, we apply the ratio of $\sigma^{\rm N^2LO}/\sigma^{\rm LO}$ from~\cite{Harlander:2003ai} for different center-of-mass energies. In the case of $b\bar{b}\to \mathcal{B}_{ps}$, the $K$-factor is about $4.1$ ($2.2$) for $m_{\mathcal{B}}=100~(400)~{\rm GeV}$, assuming the $K$-factors are similar between scalar and pseudoscalar productions. In Fig.~\ref{fig:productionrate}, we show the corresponding $m_A$ for having a $1$ fb production cross section of the bound state, which gives $300$ bound state events before passing the cuts. With a few hundreds of bound states produced at the LHC, we expect to perform bump hunts around the $2m_\chi$ invariant mass region to find out the annihilation final states.

The DLJ pair could also be produced from the off-shell pseudo-scalar mediator process $pp\to A^{*}\to Z_dZ_d$ through a one-loop process as in Fig.~\ref{fig:feydigm}, but without the internal photon lines. The differential cross section is
\begin{eqnarray}
\frac{d\sigma_{pp\to A^{*}\to Z_dZ_d}}{dm^2_{Z_dZ_d}}=\frac{m^6_{Z_dZ_d}}{8\,\pi\, s\,\Lambda_1(m^2_{Z_dZ_d})^2\,\Lambda_2(m^2_{Z_dZ_d})^2\,\left[(m^2_{Z_dZ_d}-m^2_A)^2+(\Gamma_A\,m_A)^2\right]}\\\nonumber\times\left[\displaystyle{\int^1_{m_{Z_dZ_d}^2/s}}\frac{dx}{x}f_g(x)f_{g}(\frac{m_{Z_dZ_d}^2}{x\,s})\right],
\end{eqnarray}
where the effective coupling for the dark matter loop is $\Lambda_2(q^2)^{-1}=({\alpha_\chi\,y_{\chi}}/{8\sqrt{2}\pi\,m_\chi})A_{1/2}^{A}(\tau_\chi)$ with $\tau_\chi ={q^2}/{4m_\chi^2}$. The mediator produced via the one-loop process acts as a background for the bound state search, but its signal distribution is more smooth in $m_{Z_dZ_d}$. We take into account the resolution of the invariant mass reconstruction from the DLJ search, and calculate the production rate of the dark photon pair through this process with the invariant mass within $\pm 5$~GeV from the bound state mass. 
For $m_A\gg2m_\chi$, the average event number of the one-loop production is much less than $1$ in the regions relevant for the bound state search. In particular, we find the leading-order cross section is $\sim10^{-2}~{\rm fb}$ in the same parameter space where the bound state production cross section is $\sim1~{\rm fb}$ as shown in Fig.~\ref{fig:productionrate}. Since the kinematics of dark photons and their reconstruction efficiency are similar in the both cases, the one-loop process has a negligible contribution for $m_{Z_dZ_d}\approx m_{\mathcal{B}_{ps}}\ll m_A$. 
In the search of heavier SIDM bound states, $2m_\chi$ can be closer to $m_A$ and the contribution from the one-loop process can be larger. However, in the parameter regions we are interested (Sec.~\ref{sec:results}), it is still sub-dominant compared to the assumed total background events ($10/100$).

Besides the bound state search, searching for the invariant mass peak at $m_A$ can also probe the pseudo-scalar portal model. The pseudoscalar production cross section from the gluon fusion process is
\begin{eqnarray}\label{eq:Aprod}
\sigma_{gg\to A}=\frac{G_{F}\,\alpha_{s}^2\,y_{q}^2\,m_{A}^2}{512\,\sqrt{2}\,\pi\,s}\,|\displaystyle{\sum_{q=t,b,...}}
A_{1/2}^{A}(\tau_{q})|^{2}
\times\left[\displaystyle{\int^1_{m_{A}^2/s}}\frac{dx}{x}f_g(x)f_{g}(\frac{m_{A}^2}{x\,s})\right],
\end{eqnarray}
and the decay width to a pair of dark photons is 
\begin{eqnarray}
\Gamma_{A\to Z_{d}Z_{d}}=\frac{\alpha_{\chi}^2\,y_{\chi}^2\,m_{A}^3}{256\,\pi^{3}\,m_{\chi}^{2}}|A_{1/2}^{A}(\tau_{\chi})|^{2}.
\end{eqnarray}
The direct production rate of $Z_d$ from the on-shell $A$ decay is larger than that from the SIDM bound state, but with a very different invariant mass distribution. The signal efficiency of the DLJ search based on the existing ATLAS study~\cite{ATLAS:2016jza} is up to $2\%$ for the best choice of $Z_d$ lifetime. And it sets $2\sigma$ upper bounds on the signal production to be $38~{\rm fb}$ for $e^+e^-$ final states and $6.7~{\rm fb}$ for requiring at least one pair of muons, corresponding to the lower limits on the pseudoscalar mass $m_A\sim440~(510)~{\rm GeV}$ for $m_{\chi}=50~(150)~{\rm GeV}$ with $\alpha_{\chi}=0.5$ and $y_q=y_{\chi}=1$. For $\alpha_{\chi}=0.2$ the bounds become $m_A\sim160~(385)~{\rm GeV}$ for $m_{\chi}=50~(150)~{\rm GeV}$. Depending on the assumption of $\alpha_{\chi}$, these constraints can be better or worse than the current CMS monojet limits. If the DLJ signal from the $A$ on-shell decay is observed in the future, we can determine $m_A$ from the invariant mass distribution and extract the $Z_d$ coupling $\alpha_{\chi}$ by comparing the signal rate to the dark matter bound state production. Since measuring dark matter mass has a more direct application to the SIDM parameter space, we will focus on the search of dark matter bound state in this work. If we detect the bound state signal in the future, the model predicts DLJs from the pseudoscalar decay as well.

\section{Displaced lepton jet signals}\label{sec:DLJs}
From Eq.~(\ref{eq:Zddecay}), the boosted $Z_d$ from the $\mathcal{B}_{ps}$ decay can easily have a detector size lifetime, and the decay products $e^+e^-/\mu^+\mu^-$ can easily be within a cone of small opening angle $\Delta R\ll 0.5$. Thus, the dark photons from the dark matter bound state decay can be treated as DLJs in the LHC search. The ATLAS collaboration has performed some studies on the DLJ signals~\cite{ATLAS:2016jza}. It mainly focuses on the Higgs or other heavy scalars decaying into DLJs plus missing energy. And the search can be further improved for the bound state search, such as increasing the jet $p_T$ cut and reconstructing the bound state mass. Here we study the future sensitivity of probing the SIDM bound state using the DLJ reconstruction efficiency reported in~\cite{ATLAS:2016jza} and a variation of their energy cuts.

To quantify the performance of this strategy, we simulate parton level events from the DLJ process $pp\to\mathcal{B}_{ps}\to2Z_d$ at the LHC in {\it MadGraph\_v2.6.1}~\cite{Alwall:2014hca} and estimate signal efficiencies under the energy cuts. As shown in Fig.~\ref{fig:SIDMPara} (left), we choose three $m_{Z_d}$ values and three kinetic mixing parameter for each $m_{Z_d}$. For $m_{Z_d}=20~{\rm MeV}$ and $50~{\rm MeV}$, we look for $Z_d\to e^+e^-$. For $m_{Z_d}=300~{\rm MeV}$, $Z_d$ also decays into $\mu^+\mu^-$. We show the results for both $\alpha_{\chi}=0.2$ and $0.5$, which lead to different signal rates at the LHC.

Since the trigger systems at the LHC are mostly designed for prompt decay, we use two non-conventional triggers similar to those used in Ref.~\cite{ATLAS:2016jza} in our study. 
For the $Z_d\to e^+e^-$ case, we use the \emph{CalRatio} trigger~\cite{Aad:2013txa} that requires a $Z_d$ to decay inside the HCAL. The energy deposit of $e^+e^-$ forms a jet-like object, and we require its transverse momentum sum to satisfy
\beq
\sum_{\ell=e^+e^-} p_T^{\ell}=p_T^{Z_d}\geq60~{\rm GeV}.
\eeq
For $Z_d\to\mu^+\mu^-$, we use the Scan Muon trigger~\cite{Aad:2014sca} that looks for muon signals \emph{without} the associated inner detector tracks. The momentum and separation of the two muons need to satisfy
\bea
p^{\mu_1}_T\geq20~{\rm GeV},\quad p^{\mu_2}_T\geq6~{\rm GeV},\quad \Delta R^{\mu\mu}\leq0.5.
\eea
Besides these trigger requirements, we further require the following cuts for the lepton jets ($\ell=e^{\pm}$ or $\mu^{\pm}$)
\bea\label{eq:emupt}
|\eta^{\ell}|\leq2.4,\,\, \Delta R^{\ell\ell}\leq0.5,\,\, p_T^{Z_d\to ee(\mu\mu)}\geq60\,(30)~{\rm GeV}.
\eea
Comparing to the ATLAS search~\cite{ATLAS:2016jza} that looks for $H\to2Z_d+X$, we assume a tighter cut on the azimuthal angle between two displaced lepton jets $|\Delta\phi|_{\rm LJ}>3$ for a $1\to 2$ decay process. Moreover, we do a bump hunt on the bound state using the total invariant mass of the two dark photons. We add $\slashed{E}_T\leq30~{\rm GeV}$ cut to separate our signal from the $pp\to A\to\chi\bar{\chi}+2Z_d$ dark photon radiation process.

For the dark photon signals that pass these cuts, we calculate their decay probability at different parts of the detector according to the boost factor and lifetime. The decay probability is further convoluted by a set of simplified reconstruction efficiencies adapted from Ref.~\cite{ATLAS:2016jza} in Tab.~\ref{recprob}.
\begin{table}[h]
  \center{
  \begin{tabular}{|c|c|c|}
    \hline
     & Barrel: $|\eta^{Z_d}|\leq1.5$ & Endcaps: $1.5<|\eta^{Z_d}|\leq 2.4$  \\ \hline
    $Z_d\to e^+e^-$& $L_{xy}= 2.2\textup{--}3.7~{\rm m}, P=0.7$ & $L_{z}= 4.3\textup{--}6.0~{\rm m}, P=0.5 $\\ \hline
    $Z_d\to \mu^+\mu^-$ & $L_{xy}= 2.2\textup{--}6.0~{\rm m}, P=0.5 $& $L_{z}= 4.3\textup{--}10.0~{\rm m}, P=0.6 $\\
    \hline
\end{tabular}
}
\caption{Dark photon decay distances and reconstruction efficiencies at the ATLAS barrel and endcaps.}
\label{recprob}
\end{table}

Once obtaining the signal efficiency, we present the results by calculating future reach of the pseudoscalar mass $m_A$ as a function of $m_{\chi}$, assuming different numbers of SM background events. It is challenging for us to estimate the background of long-lived particle signatures without doing detailed detector simulations. However, since the multi-jet events give the dominant background in the ATLAS DLJ search, we can estimate the background by rescaling the existing multi-jet events for a different luminosity and energy cuts.

To estimate the multi-jet background in our study, we simulate multi-jet background events in MadGraph and obtain the invariant mass distribution of the two leading jets passing the energy cuts. We first choose energy cuts similar to the ATLAS search~\cite{ATLAS:2016jza} by lowering the next to the leading jet $p_T$ cut to $30~{\rm GeV}$ and relaxing the azimuthal angle cut to $|\Delta\phi|_{\rm LJ}>0.63$. We assume the $241$ background events reported in their search, follow the multi-jet invariant mass distribution in our simulation and have the $|\Delta\phi|_{\rm LJ}$ distribution as in their Fig.~5. By comparing signal efficiencies between these relaxed cuts and our proposed cuts, we calculate the number of background events in the bound state search with $300~{\rm fb}^{-1}$ of data. When doing a bump hunt for $\mathcal{B}_{ps}$, we find about $10$ ($20$) background events for $m_{\chi}\approx 80~(150)~{\rm GeV}$, if the width of the constructed invariant mass peak is around $10~{\rm GeV}$, coming from the imperfect DLJ energy measurement. Since further improvements on the Run~3 and high luminosity study are expected due to the more control region data and better HCAL resolution~\cite{Boumediene:2017nhu,Magnan:2017exp,Pitters:2018nsl}, an even better background rejection can be achieved.

\section{Dimuon pair searches and the dark FSR process}
\label{sec:dimuon}
If the dark photon decays into muons before leaving the inner part of the tracker, the CMS dimuon search for pair production of new light bosons~\cite{CMS:2018rdr} can be used to test the $Z_d$ coupling to the dark matter particles. 
We take the existing CMS search to constrain our model parameters. Since this search requires four muons in the final states, it only applies to the case where the dark photon is relatively heavy, $m_{Zd}\geq2~m_{\mu}$. The CMS study collects dimuon pairs that originate from prompt or slightly displaced vertices within $9.8~{\rm cm}$ in the transverse plane. Muon spectroscopy information is used to trigger the events, and the higher level trigger requires a leading muon~($\mu_1$) with transverse momentum $p_T^{\mu_1}>15~{\rm GeV}$ and two more muons with $p_T^{\mu}>5~{\rm GeV}$ and $|\eta^{\mu}|<2.4$. The following cuts are imposed for the event selection
\bea\label{eq:CMScuts}
p_T^{\mu_1}>17~{\rm GeV}&,& |\eta^{\mu_1}|<0.9,\\\nonumber
p_T^{\mu}>8~{\rm GeV}&,& |\eta^{\mu}|<2.4,\\\nonumber
L_{xy}\leq9.8~{\rm cm}&,& L_{z}\leq46.5~{\rm cm}.
\eea 
The main background in the CMS search is from the $b\bar{b}$ process, and the $J/\psi$ and electroweak backgrounds are sub-dominant. In order to suppress the SM background, the search requires the difference between two dimuon invariant masses to be within a few times of the detector resolution, since they come from dark photons with the same mass. The search reports the total background number to be $9.90\pm 1.24\,({\rm stat})\pm 1.84\,({\rm syst})$, while $13$ events are observed with $35.9~{\rm fb}^{-1}$ data~\cite{CMS:2018rdr}.

With lower $p_T$ cuts and looser requirements on the decay location, we can recast the existing search to constrain the final state radiation (FSR) process for our model, $pp\to \chi\bar{\chi}+nZ_d$. Refs.~\cite{Bai:2009it,Buschmann:2015awa} suggest that the FSR signal can be used to test models with a light dark photon and one can even probe model parameters by measuring the cross section ratio of multi-dark photon radiation processes. While these studies focus on scenarios with much lower dark matter mass and higher center of mass energy compared to ours, we find the FSR can still be important for our search. We follow the discussion on parton shower in~\cite{Buschmann:2015awa} to calculate the probability of producing FSR mediators for a given energy cut, and use the FSR energy spectrum to extract the decay length information. We find the process is dominated by the production and decay $A\to \chi\bar{\chi}+nZ_d$ because of the PDF suppression at a higher center of mass energy. In addition, the probability of having $n\geq 2$ depends on the model parameters $(m_A,\,m_{\chi},\,\alpha_{\chi})$ and the energy cuts. For the parameter regions we consider in Sec.~\ref{sec:results}, the average number of $m$ from each $\chi^*\to\chi+mZ_d$ process is much smaller than one. So we focus on the $n=2$ case for getting the minimum number of $Z_d$ that is required by the DLJ and dimuon searches. We further use MadGraph simulations of $A\to \chi\bar{\chi}+nZ_d$ to obtain the additional efficiencies from the experimental cuts.

We take $m_{Z_d}=300~{\rm MeV}$ as an example in this analysis. Since the final state muons from the dark photon decay tend to be collimated, we assume the energy cut efficiency could be reproduced by requiring a leading FSR $Z_d$ with $p_T>34~{\rm GeV}$ and a second FSR $Z_d$ with $p_T>16~{\rm GeV}$. The dark photon energy distribution peaks around the energy cut value, and we use the spectrum to calculate the probability of finding both $Z_d$ particles to decay within the region required by the CMS dimuon search. It is $\sim0.01\textup{--}1\%$, depending on the assumption of the dark photon lifetime. In the further event reconstruction, we take the single muon track reconstruction efficiency to be $90\%$. Combining this reconstruction efficiency and the pseudo-rapidity cuts in Eq.~(\ref{eq:CMScuts}) leads to an efficiency of $50\%$. We estimate the number of $nZ_d$ events that could be detected in the dimuon search, and find the current result only excludes a small $(m_{\chi},\,m_A)$ region, where the dark matter production rate is high and the FSR process is intense. 

For the DJL search, the typical probability of having two FSR mediators with energy $> 60\,(30)$ GeV for the electron and muon searches is $\sim0.1\textup{--}1\%$. Since $\sigma_A\times$BR$(A\to\chi\bar{\chi})$ is below $3$ pb for the allowed model parameters and the optimal reconstruction efficiency of two mediators is $1\%$, the expected number of $\chi\bar{\chi}+2Z_d$ signals in the $3.6$ fb$^{-1}$ search~\cite{ATLAS:2016jza} is less than $1$. Thus, the current DLJ search is not sensitive to the FSR process. For the future DLJ search we proposed in Sec.~\ref{sec:DLJs} that focus on the bound state decay, the cuts $|\Delta\phi|_{\rm LJ}>3$ and $\slashed{E}_T\leq30~{\rm GeV}$ introduce an additional $10\%$ suppression of the signal events. The number of events in a $10$ GeV bin around the bound state peak is always less than $1$. Hence, we can separate the FSR signal from the bound state signal in the collider search, and the FSR signal in the dimuon search provides an additional probe to the SIDM model. We will show detailed sensitivity limits in Sec.~\ref{sec:results}.

\section{Results and discussion}
\label{sec:results}

\begin{figure}
\begin{center}
\includegraphics[height=7cm]{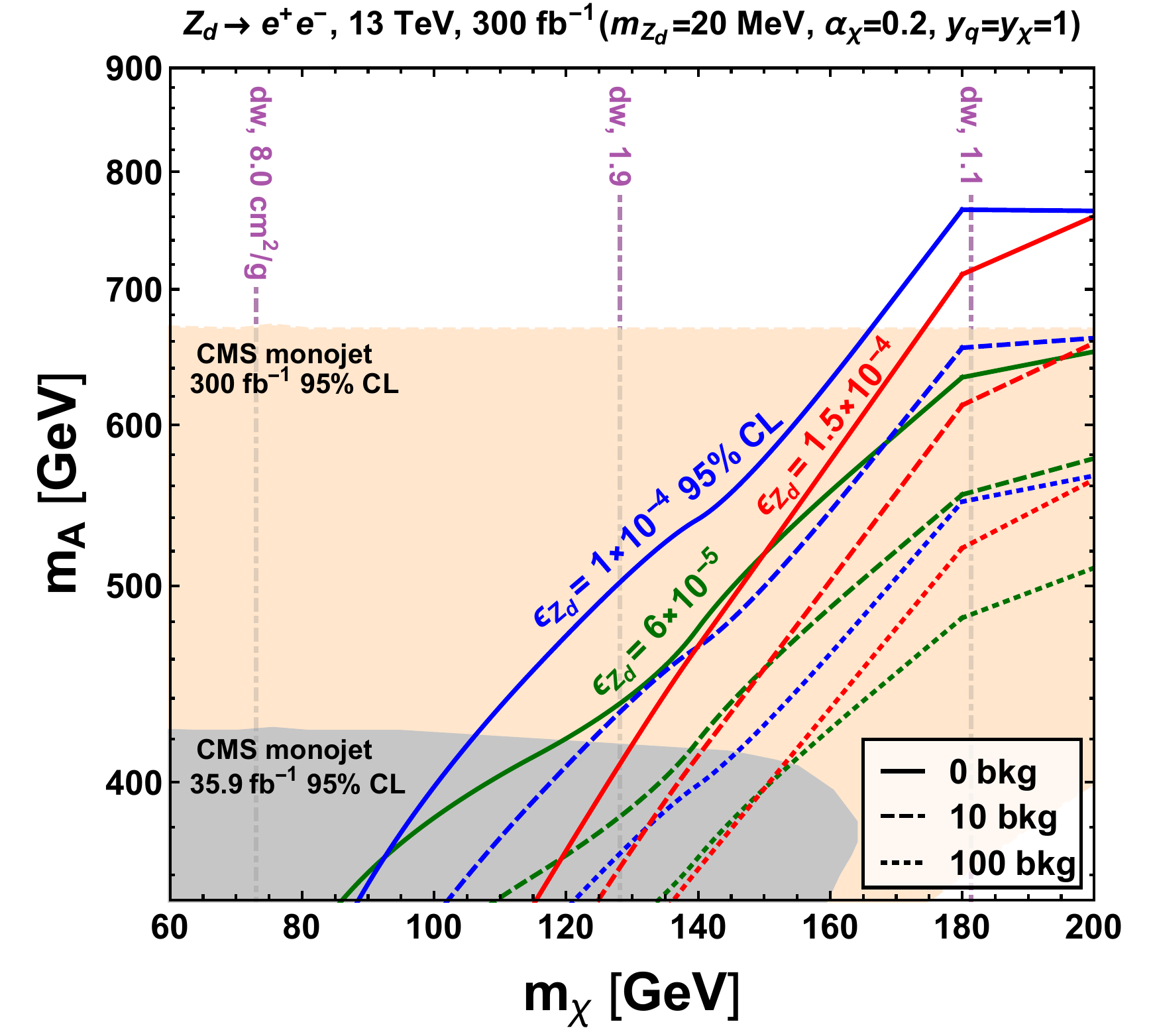}\,\,\,\includegraphics[height=7cm]{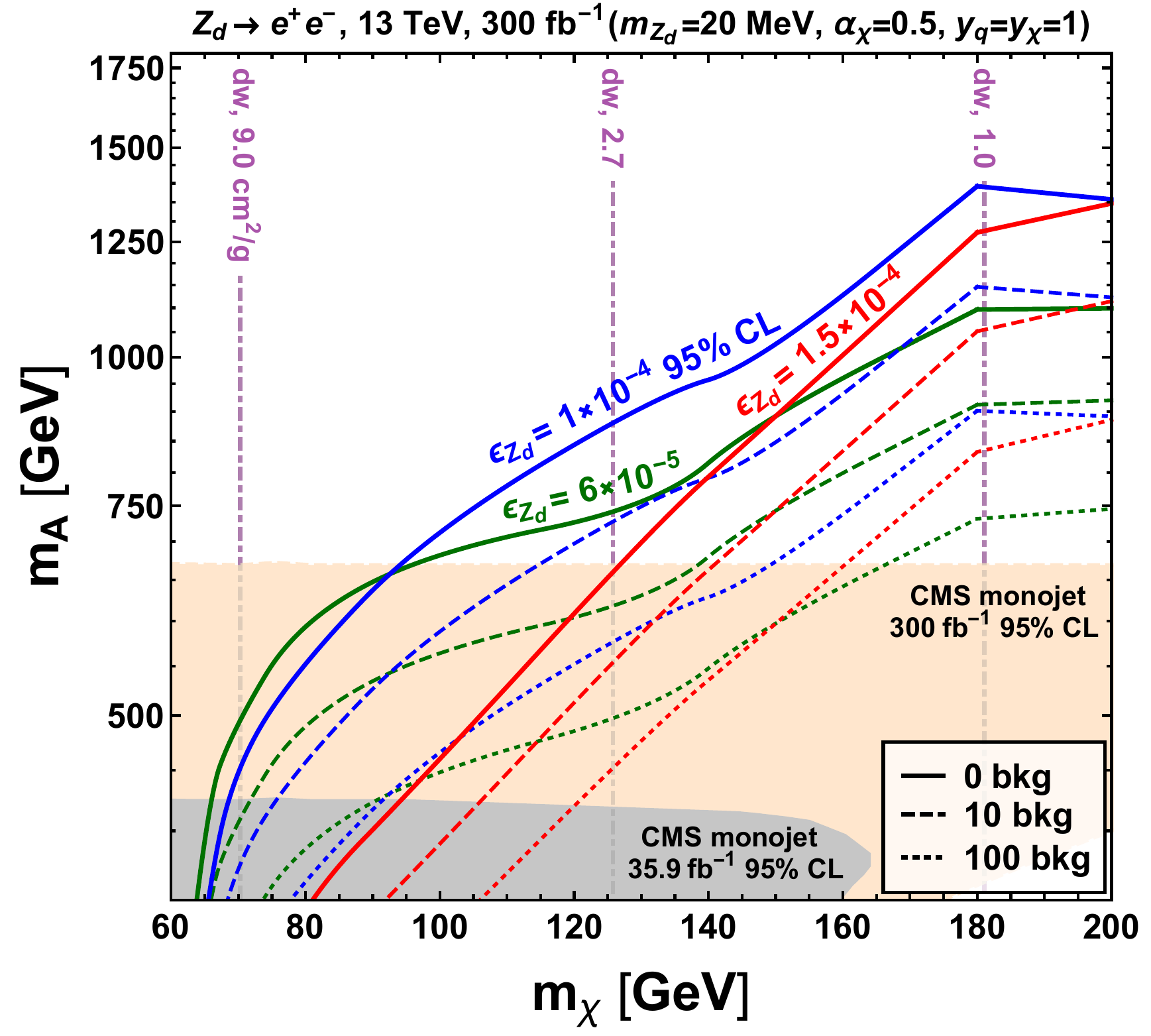}
\\\vspace{1em}
\includegraphics[height=7cm]{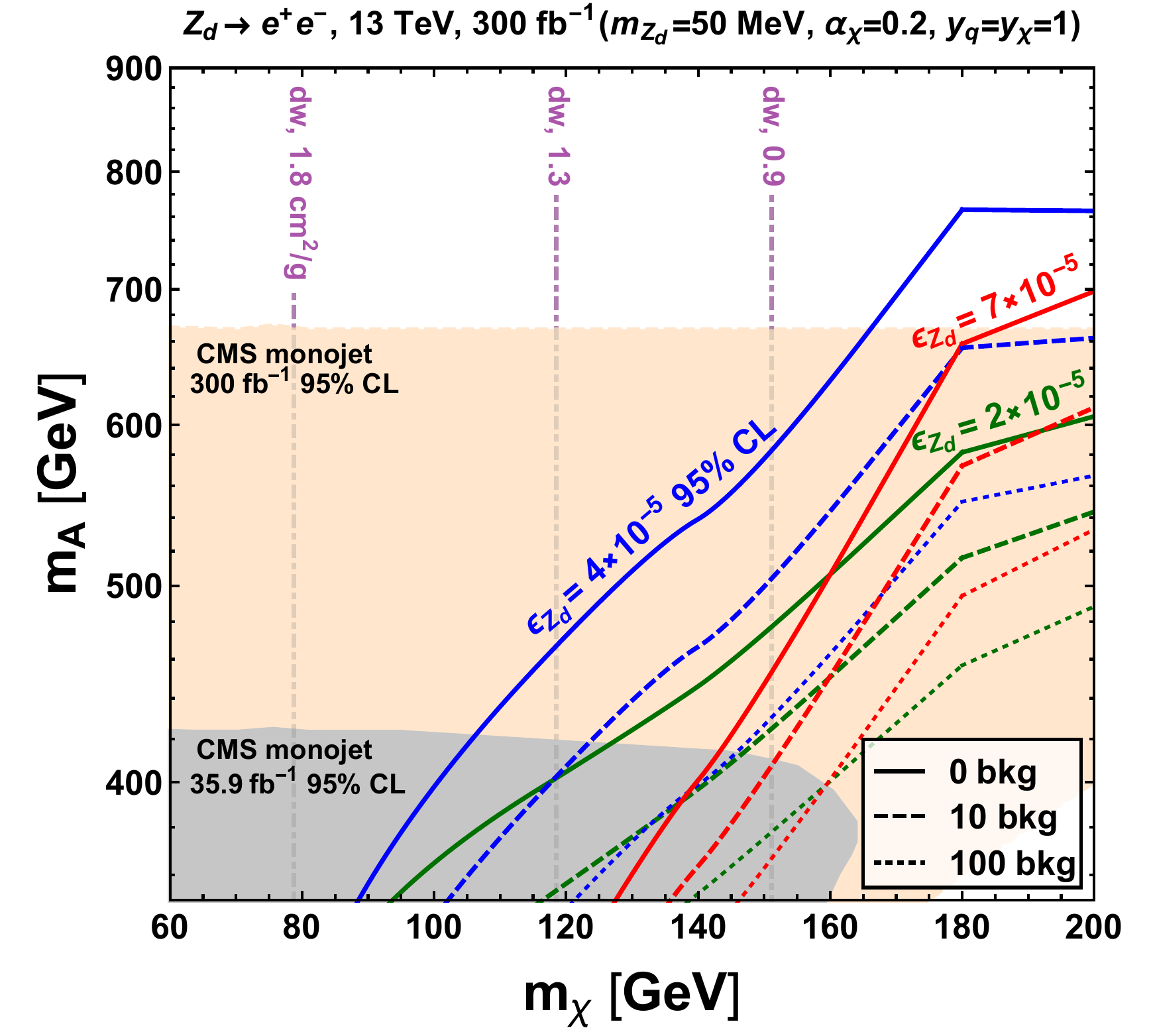}\,\,\,\includegraphics[height=7cm]{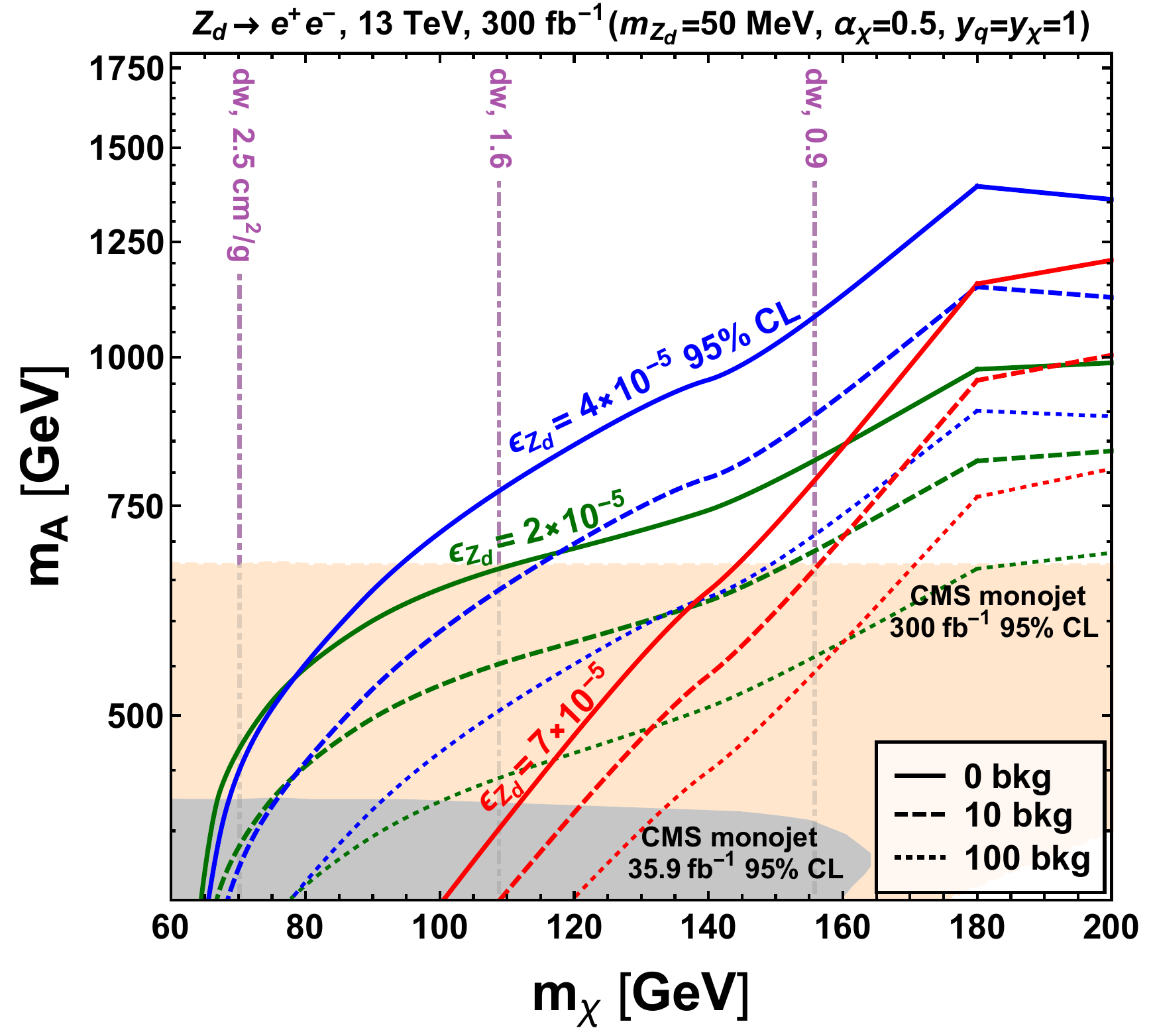}
\end{center}
\caption{Projected $95\%$ C.L. bounds of $m_A$ in $Z_d\to e^+e^-$ channel at $\sqrt{s}=13$~TeV LHC for different kinetic mixing parameters (blue, green and red). We take the dark coupling constant as $\alpha_{\chi}=0.2$ (left) and $0.5$ (right). For each panel, we assume different background events, $0$ (solid), $10$ (dashed) and $100$ (dotted). The grey region is excluded by the CMS monojet search of the pseudoscalar with $35.9$~fb$^{-1}$ of data, and the orange one indicates the projected bound. Vertical dashed-dotted lines (purple) denote the corresponding dark matter self-scattering cross section in dwarf galaxies, and $\sigma/m_{\chi}\approx1\textup{--}10~{\rm cm^2/g}$ is favored to explain the astrophysical observations.} 
\label{fig:phiee}
\end{figure}

\begin{figure}
\begin{center}
\includegraphics[height=7cm]{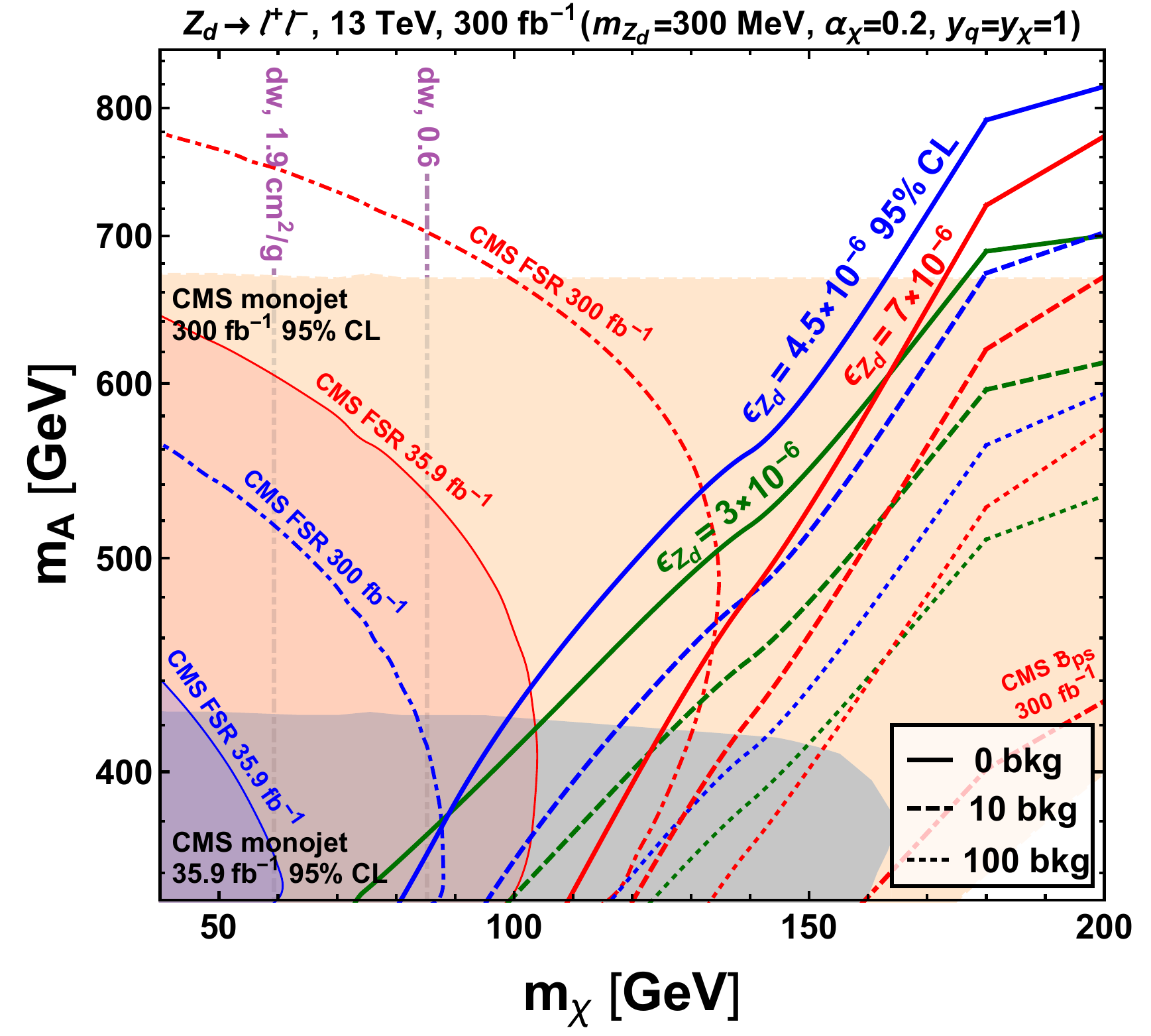}\,\,\,\includegraphics[height=7cm]{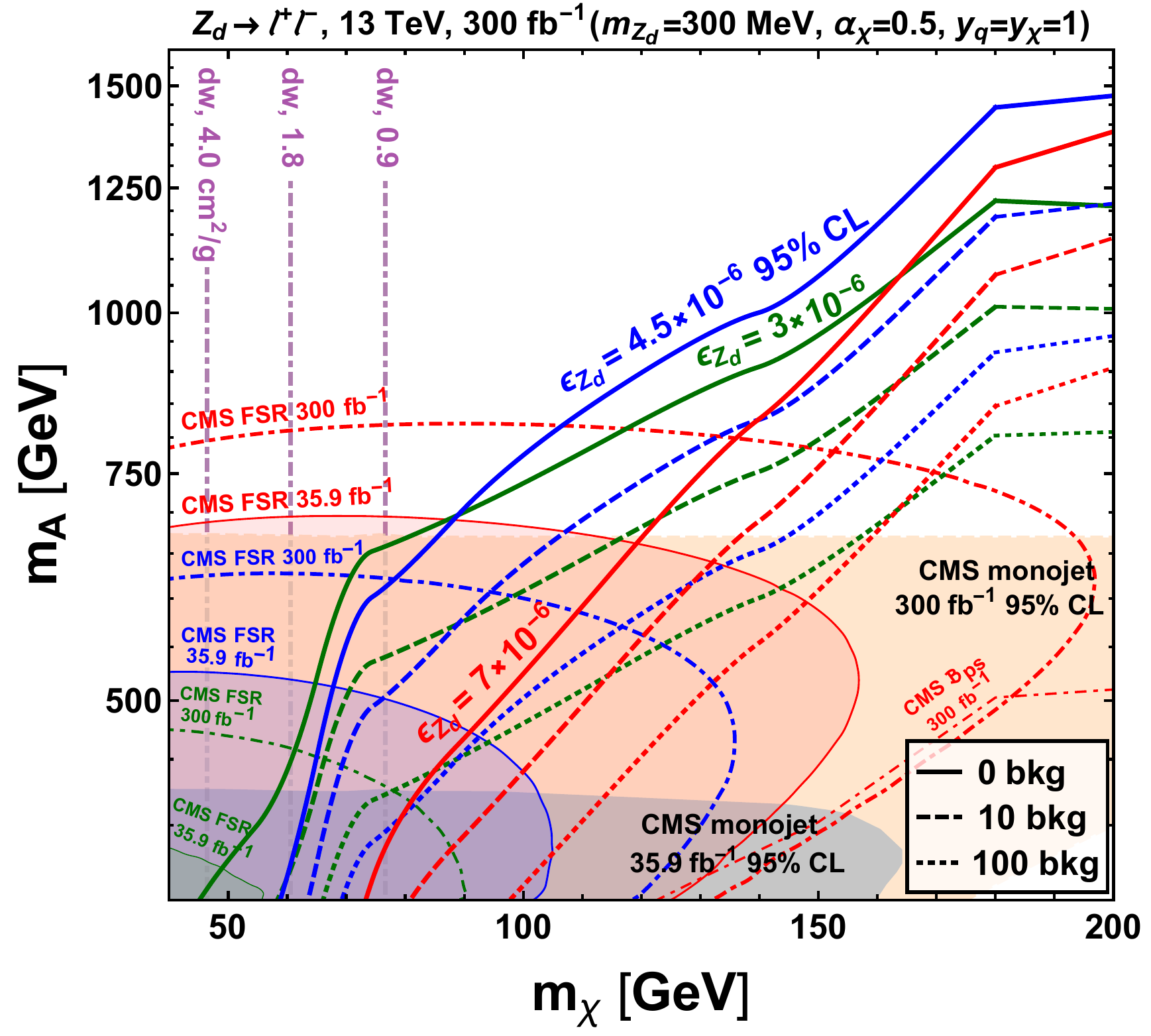}
\end{center}
\caption{Similar to Fig.~\ref{fig:phiee} but with a $300~{\rm MeV}$ dark photon and decay channels $Z_d\to e^+e^-$ and $\mu^+\mu^-$. The red, blue and green shaded regions are excluded by the CMS dimuon search for the dark photon FSR process for $\epsilon_{z_{d}}=7\times10^{-6}$, $4.5\times10^{-6}$ and $3\times10^{-6}$, respectively. The red, blue and green dashed-dotted curves on the left corner denote future sensitivity to the FSR process after LHC Run 3. The red dashed-dotted curves on the bottom right corner denote the projected CMS dimuon search sensitivity to the SIDM bound state production. }
\label{fig:phimumu}
\end{figure}

In Fig.~\ref{fig:phiee}, blue, green and red curves denote the reach of the pseudoscalar mass $m_A$ from the DLJ searches for $Z_d\to e^+e^-$. We assume a $13~{\rm TeV}$ search with $300~{\rm fb^{-1}}$ of data, and set the pseudoscalar coupling constants to be $y_q=y_{\chi}=1$, the same as in the CMS monojet study~\cite{Sirunyan:2017jix} that found a lower bound on $m_A$ (gray, $95\%$ CL). We also show the projected reach of future monojet search (orange) using the Collider Reach tool~\cite{collreach}, based on the scaling of PDF and luminosity. Fig.~\ref{fig:phimumu} shows similar results for $Z_d\to e^+e^-$ and $\mu^+\mu^-$.

We assume the sensitivity is statistically dominated and show the results with different background events, $0$ (solid), $10$ (dashed) and $100$ (dotted). As discussed in Sec.~\ref{sec:DLJs}, a rescaling of the multi-jet background from the ATLAS search~\cite{ATLAS:2016jza} may provide an estimate of the expected background in the experiment. The total background events also include small contributions from the one-loop production as discussed in Sec.~\ref{sec:darkphoton}. We find the sensitivity is close to the dashed curves ($10$ background events) for $m_{\chi}\approx 80\textup{--}150~{\rm GeV}$.  For the mass and coupling we consider, the typical signal efficiency of the bound state search is $\sim1\%$ with the suppression mainly comes from the probability of having both $Z_d$'s to decay after reaching the HCAL. The suppression from energy cuts is mild when $m_{\chi}$ is larger than the cuts. In Fig.~\ref{fig:productionrate}, the $m_A$ values correspond to $300$ events in the Run~3 data, which give few signal events after taking the searching efficiency into account. Therefore, the $m_A$ values in Fig.~\ref{fig:productionrate} are close to the reach in Fig.~\ref{fig:phiee} and~\ref{fig:phimumu} denoted as blue solid curves.

In each panel, we also show the corresponding dark matter self-scattering cross section in dwarf galaxies, assuming an attractive interaction between dark matter particles. In this case, dark matter self-scattering exhibits a resonant behavior and its peak locations depend on $m_\chi$ for fixed $\alpha_\chi$ and $m_{Z_d}$. We choose representative values of $m_\chi$ so that $\sigma/m_\chi$ is in the range of $\sim1\textup{--}10~{\rm cm^2/g}$ as favored by astrophysical observations in dwarf galaxies. We see that the DLJ search proposed in this work provide a complementary probe to the SIDM parameter space.

As discussed in the previous section, the CMS search of prompt dimuon pairs can probe the FSR and bound state signals for the SIDM model. In Fig.~\ref{fig:phimumu}, we show the exclusion regions for $Z_d$ FSR from the existing CMS dimuon search~(shaded) and the future Run 3 projection~(dashed-dotted). For kinetic mixing values of $\epsilon_{Z_d}=7\times10^{-6}$, $4.5\times10^{-6}$ and $3\times10^{-6}$, the results are shown in red, blue and green, respectively. With the $35.9~{\rm fb}^{-1}$ data, we can exclude $m_\chi$ up to about $105$~($60$)$~{\rm GeV}$ when $m_A=400$~($350$) and $\epsilon_{Z_d}=7\times10^{-6}$ ($4.5\times10^{-6}$) for $\alpha_\chi=0.2$ (left panel). And the bounds become stronger for $\alpha_\chi=0.5$ accordingly (right panel). Future improvements of the dimuon search could cover a larger parameter region, and may allow simultaneous observations of DLJ and FSR events. Meanwhile, the dimuon constrains on the bound state process is very weak. For $\epsilon_{Z_d}=7\times10^{-6}$, the largest shown in Fig.~\ref{fig:phimumu}, the future exclusion limits (right corners) with $300~{\rm fb}^{-1}$ data are much weaker than the proposed DLJ searches. The constrains are even weaker for smaller kinetic mixing values.

If the SIDM parameters are in the overlap region between the monojet and DLJ searches as shown in Figs.~\ref{fig:phiee} and~\ref{fig:phimumu}, we expect three types of signals in the Run~3 data: the monojet signal and two resonance peaks in the DLJ search -- one from the decay of dark matter bound state and the other from the on-shell decay of $A$.
Since $m_A$ is always much larger than the bound state mass $m_{\mathcal{B}}$ in the parameter region of our interest, the two peaks would be distinguishable when we perform bump hunts. We can infer the dark matter mass $m_\chi$ and the pseudoscalar mass $m_A$ from the DLJ signals, and the size of dark photon coupling $\alpha_{\chi}$ by comparing the relative production rate between Eqs.~(\ref{eq:Aprod}) and (\ref{eq:Bprod}). After the events are accumulated sufficiently enough, we can fit $m_{Z_d}\epsilon_{Z_d}$ in Eq.~(\ref{eq:Zddecay}) from the $Z_d$ decay length, and obtain $y_\chi$ by taking the inferred $y_q$, $\alpha_{\chi}$ and $m_{\chi}$ values and the decay probability as an input to calculate the bound state production rate and compare it with the observed one. Finally, we can determine $m_{Z_d}$ from astrophysical constraints on $\sigma/m_\chi$ since both $\alpha_\chi$ and $m_\chi$ are known from the collider researches, and further extract the kinetic mixing parameter $\epsilon_{Z_d}$ from the lifetime measurement. Thus, it is possible to completely fix the SIDM model parameters by combining different LHC searches and astrophysical observations if a positive signal is detected.

\section{Conclusions}
\label{sec:con}
SIDM is a well-motivated dark matter theory that solves the long-standing problems of cold dark matter on galactic scales. In many particle physics realizations of SIDM, there exists a light dark force mediator, which can lead to signals at terrestrial experiments if it couples to the SM particles. In this paper, we have constructed a simple particle physics model and studied its signals at the $13~{\rm TeV}$ LHC with an integrated luminosity of $300~{\rm fb^{-1}}$. For the model parameters favored by astrophysical observations, SIDM particles produced at the LHC can form a dark matter bound state, which further decays to lepton jets with displaced vertices, resulting in a striking signature with few SM backgrounds. Compared to the traditional monojet signal, the DLJ search can significantly improve the bound on the mass of the heavy pseudoscalar, mediating dark matter and SM fermion interactions. If a positive signal is detected in the future, we may determine the SIDM particle mass and the dark coupling constant by measuring the production cross section and invariant mass. Our results demonstrate the LHC can provide a complementary search for the self-interacting nature of dark matter.

\begin{acknowledgments}
We thank Gerardo Alvarez, Yangyang Cheng, Gail Hanson, Zhen Liu, Tao Ren, Weinan Si, Yue Zhang and Yue Zhao for useful discussions. The work of YT was supported in part by the National Science Foundation under grant PHY-1620074, and by the Maryland Center for Fundamental Physics. YT thanks the Aspen Center for Physics, which is supported by National Science Foundation grant PHY-1066293. HBY acknowledges support from U.S. Department of Energy under Grant No. de-sc0008541 and UCR Regents' Faculty Development Award.  
 \end{acknowledgments}

\bibliography{sidm}
\bibliographystyle{JHEP}

\end{document}